# Nonzero RMS Magnetoresistance Yielding Control Space Partition of $CrTe_2$ Monolayer

Chee Kian Yap[1], and Arun Kumar Singh[2]

1 Queensland University of Technology, Australia, E-mail: cheekian.yap@hdr.qut.edu.au

2 ICFAI University, Raipur, Chhattisgarh, India

**Abstract.** The study of magnetic phenomena in low-dimensional systems has largely explored after the discovery of two-dimensional (2D) magnetic materials, such as $CrI_3$ and $Cr_2Ge_2Te_6$ in 2017. These materials presents intrinsic magnetic order, overcoming the limitations predicted by the Mermin-Wagner theorem, due to magnetic crystalline anisotropy energy. Among these, $CrTe_2$, a van der Waals 2D magnet, has gather significant interest due to its in-plane anisotropic magnetoresistance (AMR) and high Curie temperature. This study investigates the magnetic field-regulated resistance of $CrTe_2$ monolayers in the context of spintronics applications. Utilizing the zigzag-ordered parameters obtained from prior simulations, we examine how external magnetic fields influence resistance states and control the ON/OFF state of nano-devices. The analysis demonstrates that specific magnetic field configurations, particularly those in the form of $(0, 0, B_z(\omega))$, which is out-of-plane directed field, gives a non-zero root mean square resistance, indicating a functional ON state. This provides a novel method for magnetically controlled current regulation in spintronic devices. The experimental results also reveal an interesting spin-flop transition in $CrTe_2$ under a z-directed magnetic field, leading to y-directional magnetization. This phenomenon, combined with the material's robust magnetic properties, positions $CrTe_2$ as a promising candidate for next-generation memory and logic devices. By advancing the understanding of magnetic field manipulation in 2D magnetic materials, this research opens new pathways in the development of energy-efficient spintronics technology.

# 1. Introduction

Most magnets known before the discovery of two-dimensional atomic-thin magnets are bulk in nature. Early research has proven magnetic order cannot be realized in one dimensional at finite temperature e.g. spin chain, though it restores order at zero temperature. Moreover, it is rigorously proved in Wagner-Mermin Theorem that long-range magnetic order is suppressed at finite temperature for lower dimensional materials that are described by isotropic Heisenberg model. The surprise discovery of 2D magnetic materials $CrI_3$, and $Cr_2Ge_2Te_6$ in 2017[2, 3] open new research avenues that has fundamental importance, such as the theory of phase transition, in addition to bringing advance in technology, such as in spintronics. The appearance of intrinsic magnetic order in 2D materials is generally attributed to magnetic crystalline anisotropy energy that endow the system with a magnon gap, thereby breaking the rotational symmetry of spins, becoming a symmetry reduced system. Recently, an attractive new class of van der Waals 2D magnet $CrTe_2$ has been found, it is reported to have in-plane magnetoresistance and high Curie temperature[4]. The order parameters are in silico and ab-initio determined. Experimentally, it has reported an intriguing effect that the z-directed magnetic field on the monolayer induces y directional magnetization, a spin-flop transition to rotate the moments parallel to the z direction[5], furthermore, the work of Yan Feng et. al. shown that the out-of-plane z magnetic field has a stronger effect to magnetoresistance than in-plane directed magnetic field[6]. In this work, we used the obtained a zigzag-ordered parameters[7] and calculate the response of magnetic field regulated resistance, which in turns, control the ON/Off state of the tiny device. We find that certain points in the control space partition of external magnetic fields in the form of $(0, 0, B_z^*(\omega))$ yield non-zero root mean square of resistance which can signifies a turn ON state, demonstrating ability to control the current signals by magnetism via the control strategy.

Fitting more electronic transistors into smaller areas in the miniaturization process is an approach to achieving low energy consumption and high-density logic and memory, and it can scale up the performance of the chip as well. In such applications, the mechanism is based on charge carrier transport. In contrast, spintronics exploit spin degrees of freedom of electrons in devices, such as in magnetoresistive devices, and it consumes less energy in manipulation of information, and can achieve denser implementation of memory and logics, despite having slower performance compared to charge transport.

The material CrTe₂ monolayer was selected in this study because it was synthesized and reported as having in-plane AMR (Anisotropy Magnetic Resistance)[4]. CrTe₂ belongs to a class of Transition Metal Dichalcogenide (TMD) materials in the $1T$ phase that can be synthesized through chemical vapor deposition or mechanical exfoliation. It has high Curie temperature and long-range exchange interaction[1] and chemical stability in air. It is a ferromagnet, in its bulk phase or with thickness of multiple layers. It possess an out-of-plane easy-axis orientation in a minimum of four monolayers[8]. However, in its monolayer limit, it is a room-temperature anti-ferromagnet (AFM) with the in-plane AMR of -0.6% at 300 K[4] where the negative in AMR indicates the vertical directed magnetism has a higher resistance, the in-plane magnetism is shown in Fig. 1. The AMR is given as $\frac{R_\parallel - R_\perp}{R_\perp}$%. The AFM-Zigzag configuration of the monolayer magnet is reported to be the most stable[9], and the FM configuration has a higher energy than the other two AFM configurations of sAFM-ABAB and sAFM-AABB. The collinear spin-polarized band calculations confirmed 1T-CrTe₂ is a metallic state in sAB configuration, the symmetry of the density of states of spin up and down further implied an AFM magnetic state, as shown in Fig. 2. The Cr atom d-orbitals resolved band structure in Fig. 3 indicate that all 5d-orbitals of Cr atoms participate in filling up the energy of the total system, with $d_z$ contributed more significantly near X symmetrical point in reciprocal space.

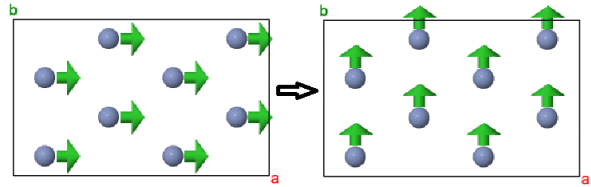

**Figure 1.** Showing in-plane magnetization, with $R_\perp > R_\parallel$ .

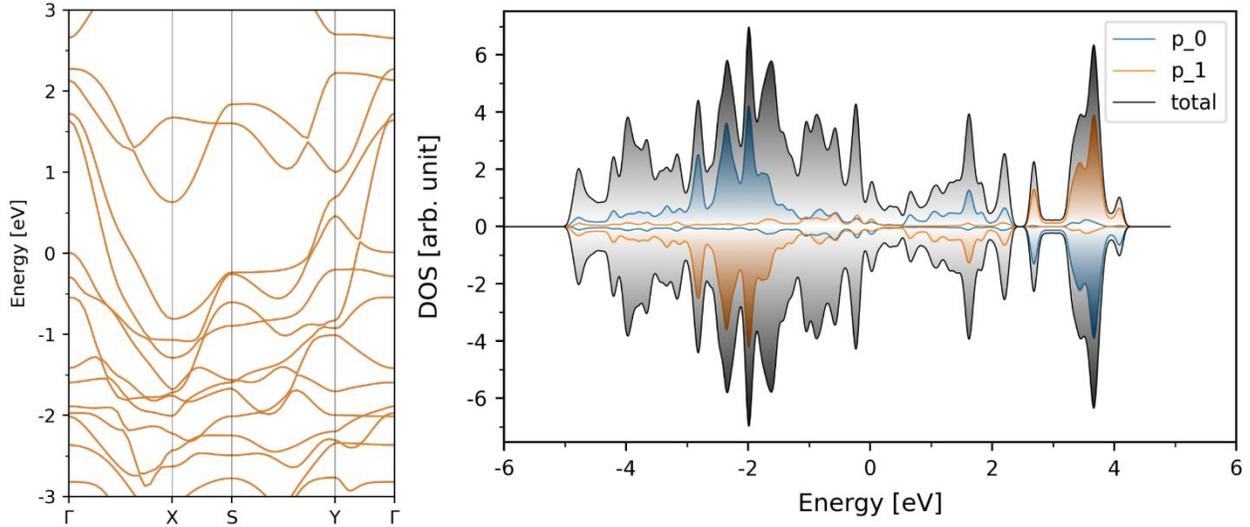

**Figure 2.** (a) Band structure of sAFM-ABAB magnetic state showing overlapping of spin majority and spin minority bands forming singlet states in AFM. There is no gap near the Fermi level, and the band intersection with the Fermi level indicates a metallic electronic state. (b) Density of States of 1T-CrTe2 sAFM-ABAB.

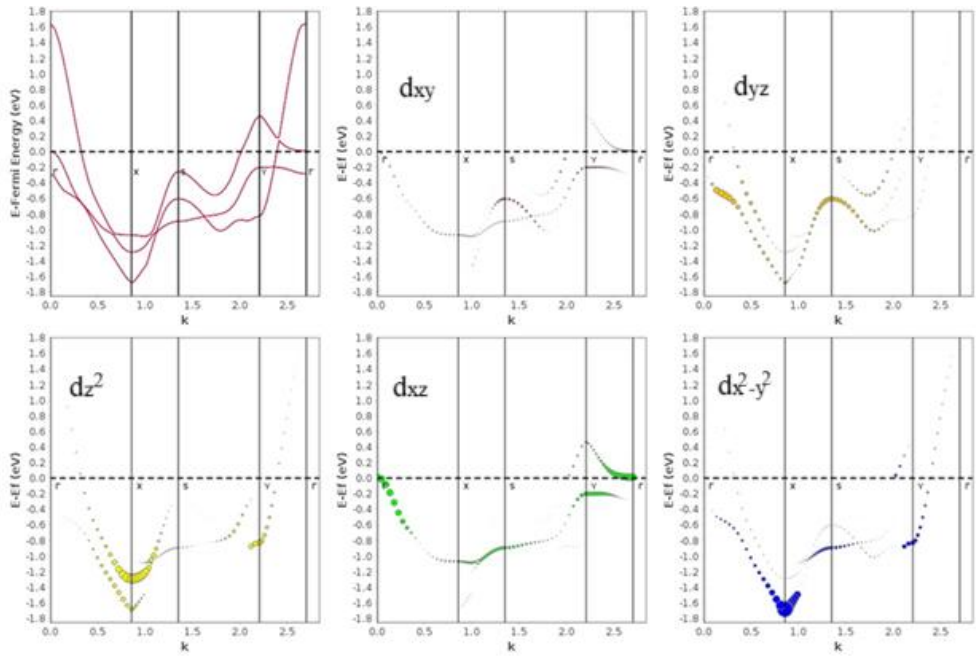

**Figure 3.** Cr atom d-orbitals resolved band structure for the three bands closest to the Fermi level.

## 2. Model

For a general magnetic system described by the isotropic XXX Heisenberg model up to four nearest neighbors for both exchange and DM interactions, the Hamiltonian of the system is written as:

$$H = -\sum_{\langle i,j \rangle}^{4} J_{ij}\, \mathbf{S}_i \cdot \mathbf{S}_j + \sum_{\langle i,j \rangle}^{4} \mathbf{D}_{ij} \cdot (\mathbf{S}_i \times \mathbf{S}_j) - \sum_i K(\mathbf{S}_i)^2 - \sum_i \mathbf{h} \cdot \mathbf{S}_i \quad (1)$$

$$S_i^\alpha = I \otimes I \ldots \otimes \sigma^\alpha \otimes \ldots I \otimes I \quad (2)$$

$$\sigma^x = \begin{bmatrix} 0 & 1 \\ 1 & 0 \end{bmatrix}, \sigma^y = \begin{bmatrix} 0 & -i \\ i & 0 \end{bmatrix}, \sigma^z = \begin{bmatrix} 1 & 0 \\ 0 & -1 \end{bmatrix} \quad (3)$$

where $J > 0$ is the FM exchange coupling, $D_{ij}$ is the Dzyaloshinskii-Moriya interaction (DMI), $\sigma_i^\alpha$ are tensor product block with $\sigma^x$ are the Pauli matrices, $K$ is the single ion anisotropy coefficient, $h$ are external magnetic fields, and $I$ is a two-dimensional identity matrix.

For a total of $N$ magnetic atoms, the output, expectation of magnetization (order parameter) along $\alpha$, with inputs $h = (Bx, By, Bz)$ and arbitrary initial conditions, is expressed as:

$$\langle m_\alpha \rangle = \frac{\sum_i^N \sigma_i^\alpha}{N}. \quad (4)$$

The resistance is written as:

$$R(\theta) = R_\perp + (R_\parallel - R_\perp)\cos^2\theta, \quad (5)$$

where $R_\perp$ is the resistance of magnetization normal to the flow of current, and $R_\parallel$ is the resistance parallel to the flow of current with $R_\perp > R_\parallel$, for simplicity we set $R_\perp=1$ and $R_\parallel=0$. $\theta$ is the in-plane angle between magnetization of y and x and is expressed as:

$$\theta = tan^{-1}\frac{\langle m_y \rangle}{\langle m_x \rangle}. \quad (6)$$

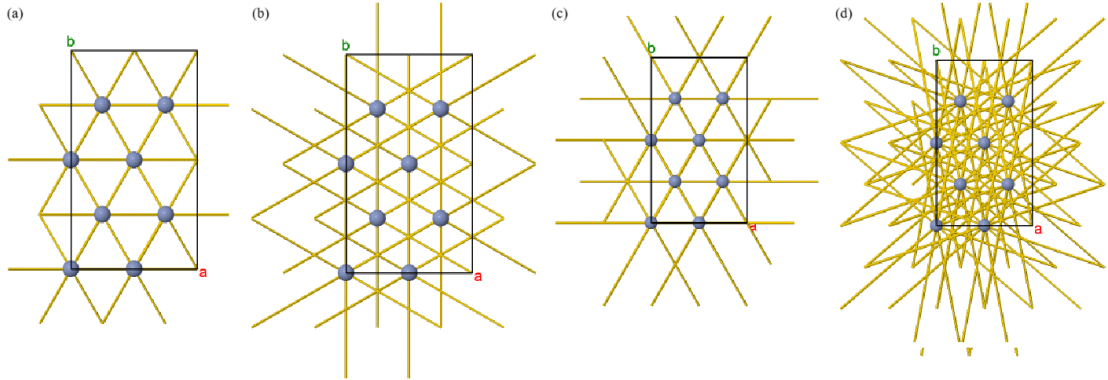

**Figure 4** Interactions of (a) 1st-nearest neighbors (b) 2nd-nearest neighbors (c) 3rd-nearest neighbors (d) 4th-nearest neighbors. Atoms in unit cell images are not shown for clarity.

## 3. Results and Discussions

Fed with parameters from ab-initio calculations[7], we set $J_1$, $J_2$, $J_3$, and $J_4$ with values -13.4, -14.9, -7.6, and 2.1 respectively, the fifth distance exceed 1 nm, therefore it is not included. We also set $K$, $D_1$, $D_2$, $D_3$, and $D_4$ with values 1.2, 0.0, 0.2, 0.0, and 0.3 respectively. The simulation cell is an enlarged orthogonal cell containing 8 Cr atoms, the 8 Cr atoms are allowed to interact with its images outside of the cell. Periodic boundary conditions are imposed on magnetic moments of all the atoms. From the results in Fig. 5 below, when the frequency of magnetic pulse is increased, we see more movements on the spins trying to align, all the calculations have root mean square of resistance greater than 0.5. It achieves greater control with a higher magnitude of magnetic pulse, where the fluctuation of spins is captured and contained within the pulse waveform.

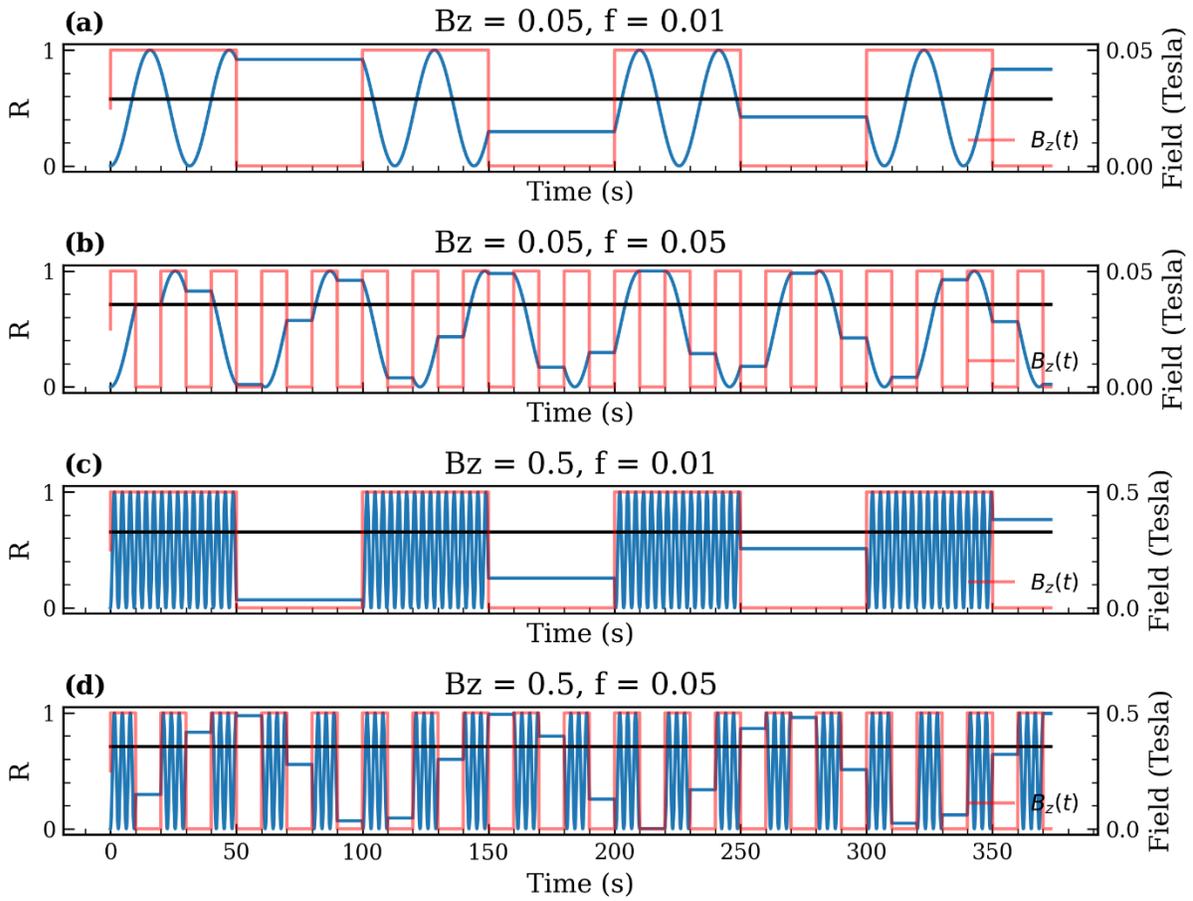

**Figure 5.** Response of magnetization-translated resistance by magnetic pulse. The black line indicates RMS, $B_z$ is the magnitude of z-directed field and $f$ is the pulse frequency.

## 4. Conclusion

The discovery of two-dimensional magnetic materials, particularly $CrI_3$ and $Cr_2Ge_2Te_6$, has deepen the exploration on the field of magnetism. The introduction of these 2D magnets has brought advancements in both fundamental physics and technological applications, especially in the realm of spintronics. In this study, $CrTe_2$ monolayers have demonstrated unique magnetic properties such as anisotropic magnetoresistance (AMR) and a remarkable spin-flop transition. The interplay of magnetic interactions, external magnetic fields and the resulting resistance behaviour has shown promising control over electronic states. The applications can lead to the development of next-generation memory devices and logic circuits with low energy consumption. The use of the Heisenberg model and ab-initio calculations has provided valuable insights into the behaviour of $CrTe_2$. The calculated resistance and magnetoresistance behaviours in certain partition of control space, underline the potential of this material in practical applications.